\documentclass[12pt,english,aps,prl,twocolumn,showpacs,superscriptaddress,groupedaddress,footinbib,reprint,noshowpacs]{revtex4-1}
\usepackage{color}
\usepackage{graphicx}
\usepackage{amsmath}
\usepackage{comment}
\usepackage{amssymb}

\usepackage{subfigure}

\begin{document}

\title{Controlling polyelectrolyte adsorption onto carbon nanotubes by tuning ion-image interactions} 

\author{Alpha A. Lee}
\email{aal44@cam.ac.uk}
\affiliation{Cavendish Laboratory, University of Cambridge, Cambridge CB3 0HE, UK}
\altaffiliation{AAL and SVK contributed equally to this work.}
\author{Sarah V. Kostinski}
\email{sarah.kostinski@mail.huji.ac.il}
\altaffiliation{AAL and SVK contributed equally to this work.}
\author{Michael P. Brenner}
\affiliation{School of Engineering and Applied Sciences, Harvard University, Cambridge, MA 02138, USA.}

\begin{abstract}
Understanding and controlling polyelectrolyte adsorption onto carbon nanotubes is a fundamental challenge in nanotechology. Polyelectrolytes have been shown to stabilise nanotube suspensions through adsorbing onto the nanotube surface, and polyelectrolyte-coated nanotubes are emerging as building blocks for complex and addressable self-assembly. The conventional wisdom suggests that polyelectrolyte adsorption onto nanotubes is driven by specific chemical or van der Waals interactions. We develop a simple mean-field model and show that ion-image attraction is a significant effect for adsorption onto conducting nanotubes at low salt concentrations.  Our theory suggests a simple strategy to selectively and reversibly functionalize carbon nanotubes based on their electronic structure which in turn modifies the ion-image attraction. 
\end{abstract}

\makeatother
\maketitle 

\section{Introduction}

Carbon nanotubes (CNTs) are one of the most important and ubiquitous building blocks in nano devices \cite{schnorr2010emerging,smalley2003carbon,zhang2013road}. Their plethora of applications range from transistors \cite{tans1998room,martel1998single} and electrical wires \cite{dekker1999carbon,kurzepa2014replacing} to probes for in vitro imaging of biological systems \cite{fakhri2014high,hong2015carbon,battle2016broken}. Exploiting the interaction between polyelectrolytes and CNTs is a recurring theme in many of those applications: CNTs can be dispersed in solution via polyelectrolyte adsorption \cite{liu2007debundling,deria2010phase,samanta2014conjugated} and conducting CNTs can be separated from the semiconducting CNTs \cite{zheng2003structure,zheng2003dna,strano2004understanding}, although how the electronic structure of CNTs affects adsorption is hitherto less clear. Polyelectrolyte-nanotube complexes can themselves be used as chemical sensors \cite{staii2005dna}, electro-optical materials \cite{deria2013single}, as well as building blocks for complex self-assembled structure \cite{maune2010self} such as plasmonic metamaterials \cite{tan2011building}. Moreover, understanding the interaction of CNTs with biopolymers, which are often charged, is important when assessing the in vitro toxicity of CNTs \cite{madani2013concise}. 

Previous theoretical studies and simulations explain polyelectrolyte adsorption onto CNTs in terms of specific chemical interactions, such as $\pi-\pi$ stacking between aromatic monomers (e.g. nucleobases in the case of DNA) and the nanotube surface \cite{gao2004simulation,zhao2007simulation,johnson2008probing,zhao2011self,roxbury2012molecular,iliafar2014interaction}. This microscopic picture is deduced from classical molecular dynamics simulations of CNT-polyelectrolyte interactions which neglects the electronic structure of the CNT, perhaps because the development of efficient simulation techniques to account for metallic surfaces is still ongoing \cite{siepmann1995influence,reed2007electrochemical,raghunathan2007self,vatamanu2017application}, and use Lennard-Jones potentials to model CNT-polyelectrolyte interactions. Those pioneering works show that specific chemical affinity is a sufficient condition for adsorption. Nonetheless, they raise the broader questions of whether adsorption is unique to polyelectrolytes with significant nonelectrostatic interactions and what role does the electronic structure of CNTs play in driving adsorption. 

In this paper, we argue that there is another significant effect driving adsorption -- the ion-image attraction. The ion-image attraction arises as external charges polarise electrons on the surface of a conductor. This induces an equal and opposite surface charge which lowers the self-energy of ions near conducting boundaries, thus attracts ions closer to the surface of a conductor \cite{jackson2007classical}. Unlike specific chemical interactions, the ion-image attraction is a universal feature between ions and electrical conductors, and its strength and range depends on the electronic structure of the material \cite{pastewka2011charge}. An emerging body of simulations using constant potential surfaces have already demonstrated the importance of including ion-image interactions to understand the physics of ions near interfaces, such as the arrangement of ions near metallic electrodes \cite{tazi2010potential,limmer2013charge}, in carbon nanopores \cite{wu2011complex,merlet2012molecular,merlet2012simulating,wu2012voltage}, as well as polyelectrolyte adsorption onto a planar surface with a  low dielectric constant where the ion-image interaction is repulsive \cite{netz1999adsorption}. Moreover, recent experiments directly revealed the effect of the electronic structure of the substrate on hydrodynamic flows in nanotubes \cite{secchi2016massive,majumder2017flows} and phase behaviour of electrolytes under nanoconfinement \cite{comtet2017nanoscale}. However, the role of ion-image interaction in polyelectrolyte adsorption onto CNTs remains unexplored. 

We will first introduce a mean-field theory for polyelectrolyte adsorption, and show that ion-image interactions can dominate over van der Waals interactions at low salt concentrations. We will then map the adsorption phase diagram, demonstrating how salt concentration could be tuned to allow selective adsorption of polyelectrolytes depending on the electronic structure of the CNT. The goal of our paper is to present a physical picture of how ion--image interactions could be a simple handle to control polyelectrolyte adsorption onto CNT, via estimates of the van der Waals and ion-image energies. As such, a simplified but analytically tractable model will be used. 

\section{Theory}

We model the polyelectrolyte as a polymer with linear charge density $\rho$ with $N$ freely hinged links of Kuhn length $l$ each. The CNT exerts a potential $U(\mathbf{r})$ on each segment. In the continuum limit, the probability density $G(\mathbf{r},\mathbf{r}_0, L)$ (Green's function) with one end point located at $\mathbf{r}$ and the other end point located at $\mathbf{r}_0$ is given by the equation \cite{edwards1965statistical,lifshitz1969some,wiegel1977adsorption,doi1988theory,von1994adsorption,winkler2006critical,cherstvy2011polyelectrolyte}
\begin{equation}
\left[ \frac{\partial}{\partial N} - \frac{l^2}{6} \Delta_{\mathbf{r}} + \beta U(\mathbf{r}) \right] G(\mathbf{r}, \mathbf{r}_0, L) = \delta(\mathbf{r}-\mathbf{r}_0) \delta(N). 
\label{Doi-Edwards}
\end{equation} 
where $\beta = k_B T$. The probability of finding the polymer approaches zero at infinity and at a hard wall, thus 
\begin{equation}
\lim_{|\mathbf{r}|\rightarrow \infty} G(\mathbf{r}) =0, \; \mathrm{and} \; G(\mathbf{r}_s) = 0. 
\end{equation}
where $\mathbf{r}_s$ is the location of the CNT surface. $G(\mathbf{r},\mathbf{r}_0, L)$ can be expressed in terms of a bilinear expansion 
\begin{equation} 
G(\mathbf{r},\mathbf{r}_0, L) = \sum_{n=0}^{\infty} f_n(\mathbf{r})f^{*}_n(\mathbf{r}_0) e^{-\lambda_n N}, 
\label{bilinear}
\end{equation}
where $f_n(\mathbf{r})$ and $\lambda_n $ are given by the eigenvalue problem
\begin{equation}
\left[- \frac{l^2}{6} \Delta_{\mathbf{r}} + \beta U(\mathbf{r}) \right] f_n(\mathbf{r}) = \lambda_n f_n(\mathbf{r}). 
\label{eval_problem}
\end{equation}
In the limit of a long polymer chain, $N \gg 1$, the lowest eigenvalue dominates Equation (\ref{bilinear}) and $G(\mathbf{r},\mathbf{r}_0, L) \approx f_0(\mathbf{r})f_0(\mathbf{r}_0) e^{-\lambda_0 N}$. Therefore, the polymer is adsorbed to the surface if and only if a bound state exists, which corresponds to $\lambda_0<0$. 

To compute the eigenvalues of Equation (\ref{eval_problem}), we need to estimate the polyelectrolyte-CNT interaction potential $U(\mathbf{r})$ and the persistence length $l$. First we will consider the ion-image contribution ($U_{ii}(\mathbf{r})$) to $U(\mathbf{r})$ and later we will estimate the strength of the van der Waals contribution. We ignore specific chemical interactions as our goal is to estimate how ion-image interactions, which is an universal feature of charges near conducting interfaces, could drive adsorption. $U_{ii}(\mathbf{r})$ can be estimated by considering a point charge $q$ of the polyelectrolyte located at distance ${\bf r_q}$, where $r=0$ defines the axis of the CNT which we model as an infinitely long cylinder with radius $a$.  Assuming an uniform dielectric constant $\varepsilon$, within the Debye-H\"{u}ckel approximation the potential $\phi$ satisfies 
\begin{equation}
\nabla^2 \phi - \kappa_D^2 \phi= - \frac{q}{\varepsilon} \; \delta(\mathbf{r} - \mathbf{r_q}),
\label{BVP}
\end{equation}
where $\kappa_D = \sqrt{8 \pi l_B c}$ is the inverse Debye length, with $l_B = e^2/(4 \pi \epsilon k_B T)$ the Bjerrum length and $c$ the salt concentration. The electronic structure of the CNT enters into the electrostatic boundary value problem via the metallicity (also known in the literature as the quantum capacitance \cite{luryi1988quantum}). The metallicity arises as one of the electrostatic boundary conditions: the electric field inside and outside the CNT is related to the induced charge $\rho_{\mathrm{ind}}$ via Gauss' law
\begin{equation} 
\varepsilon [\mathbf{n} \cdot \nabla \phi] = -\frac{\rho_{\mathrm{ind}}}{2 \pi a} = \frac{C_q \phi_s}{2 \pi a} ,
\label{quantum_cap}
\end{equation} 
where $\left[\; \cdot \; \right]$ denotes the jump across the CNT surface, $\mathbf{n}$ is the unit vector normal to the surface (defined to point in the direction outside of the pore). To arrive at the second equality, we assumed linear response, thus the induced charge is proportional to the potential at the surface, $ \phi_s$, with the constant of proportionality $C_q$ being the metallicity \cite{vella2016quantum}. The classical boundary condition for an ideal metal, $\phi_s =0$, is recovered for $C_q = \infty$. Solution of Equation (\ref{BVP}) with boundary condition (\ref{quantum_cap}) and the conditions of global continuity and asymptotic decay of the potential \cite{jackson2007classical} can be found via standard Fourier transform techniques. 

The ion-image interaction arises because the presence of a metallic boundary lowers the self-energy of ions. The self-energy of ions is lower if they are closer to the metallic boundary, thus the ions experience a body force. The self-energy is related to the potential via
\begin{equation}
E_{\mathrm{self}}(\mathbf{r}_q)  = q \lim_{\mathbf{r} \rightarrow \mathbf{r}_q} \left[ \phi(\mathbf{r},\mathbf{r}_q;C_q) - \phi(\mathbf{r},\mathbf{r}_q; 0) \right]. 
\label{self_energy}
\end{equation} 
Substituting the solution of Equation (\ref{BVP}) into (\ref{self_energy}), and noting that the polyelectrolyte is continuous and assumed to be homogeneously charged, we arrive at the ion-image interaction energy density 
\begin{widetext}
\begin{equation}
\beta U_{ii} (R) = -\frac{l_B}{\pi a} \left( \frac{\rho l}{e}\right)^2 \sum_{m=-\infty}^{\infty}  \int_{-\infty}^{\infty} \mathrm{d}x \! \left[ \frac{C \, I_m^2 \left(\sqrt{x^2+(\kappa_D a)^2} \right) K_m^2 \left(\sqrt{x^2+(\kappa_D a)^2} R \right)}{1+C I_m\left(\sqrt{x^2+(\kappa_D a)^2} \right) K_m\left(\sqrt{x^2+(\kappa_D a)^2} \right)} \right], 
\label{ion-image}
\end{equation} 
\end{widetext}
where we have introduced dimensionless quantities $C = C_q/(2 \pi \varepsilon)$, $R= r/a$, and $K_m$ and $I_m$ are the $m^{\mathrm{th}}$ modified Bessel function of the first and second kind.  
 
To close the problem, we now turn to estimate the persistence length $l$, which accounts for the mechanical stiffness of the polymer chains as well as the intrapolymer electrostatic interactions. For a worm-like chain polyelectrolyte, the Odijk-Skolnick-Fixman theory gives the simple expression \cite{odijk1977polyelectrolytes,skolnick1977electrostatic}
\begin{equation}
l = l_0 + \frac{l_B}{4 A^2 \kappa_D^2}, 
\label{OSF} 
\end{equation} 
where $l_0$ is the bare persistence length, and $A$ the distance between elementary charges along the chain ($l_B$ in the case of Manning condensation). %\red{Check the OSF formula -- Eq (\ref{OSF}) implies a diverging persistence length in the zero ion concentration limit}  (OSF is only valid for length scales above the crossover length $\frac{1}{\kappa} \sqrt{\frac{l_0}{l_0+l_B/(4 A^2 \kappa^2)}}$, e.g. see B.-Y. Ha and D. Thirumalai, ``Electrostatic Persistence Length of a Polyelectrolyte Chain,'' Macromolecules 1996,28, 577-581.) 
 Equation (\ref{OSF}) contains the salient physics that like-charge repulsion between charge groups on the polyelectrolyte increases stiffness, thus the persistence length is a decreasing function of the Debye length. 

The system of equations, Equations (\ref{eval_problem}), (\ref{ion-image}) and (\ref{OSF}), is the salient result of this paper. Those equations relate the sign of the ground state eigenvalue, $\lambda_0$, to the dimensionless parameters $\kappa a$, $C_q/2\pi \varepsilon$, and $l_B \rho l / \pi a e$. %\red{what are the dimensionless parameters??}. 
We consider typical parameters of $C_q \approx 2 \times 10^{-10} \; \mathrm{F/m}$ \cite{rotkin2006applied,pak2013relative}, $a \approx 0.4 \; \mathrm{nm}$, $c \approx 1 \; \mathrm{mM}$, and $\epsilon = 80$, $l_0 = 50 \; \mathrm{nm}$ \cite{baumann1997ionic} (parameters for double stranded DNA are used as a typical estimate for polyelectrolytes), with effective linear charge density $\rho = 1/l_B$ to account for Manning counterion condensation. 

\section{Results}

\begin{figure}
\centering
\subfigure[]{
\includegraphics[height=0.29\textwidth]{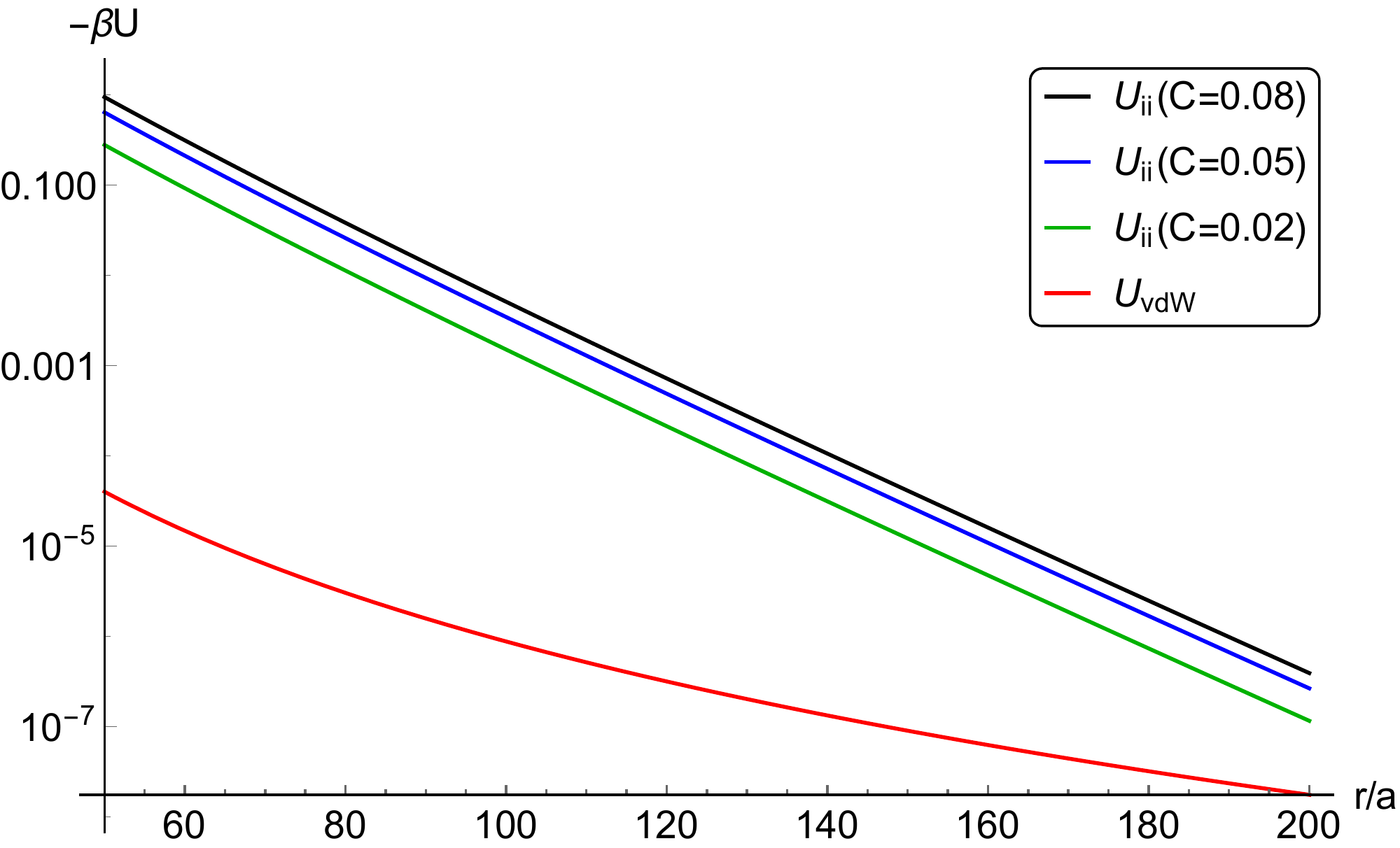}}
\subfigure[]{
\includegraphics[height=0.29\textwidth]{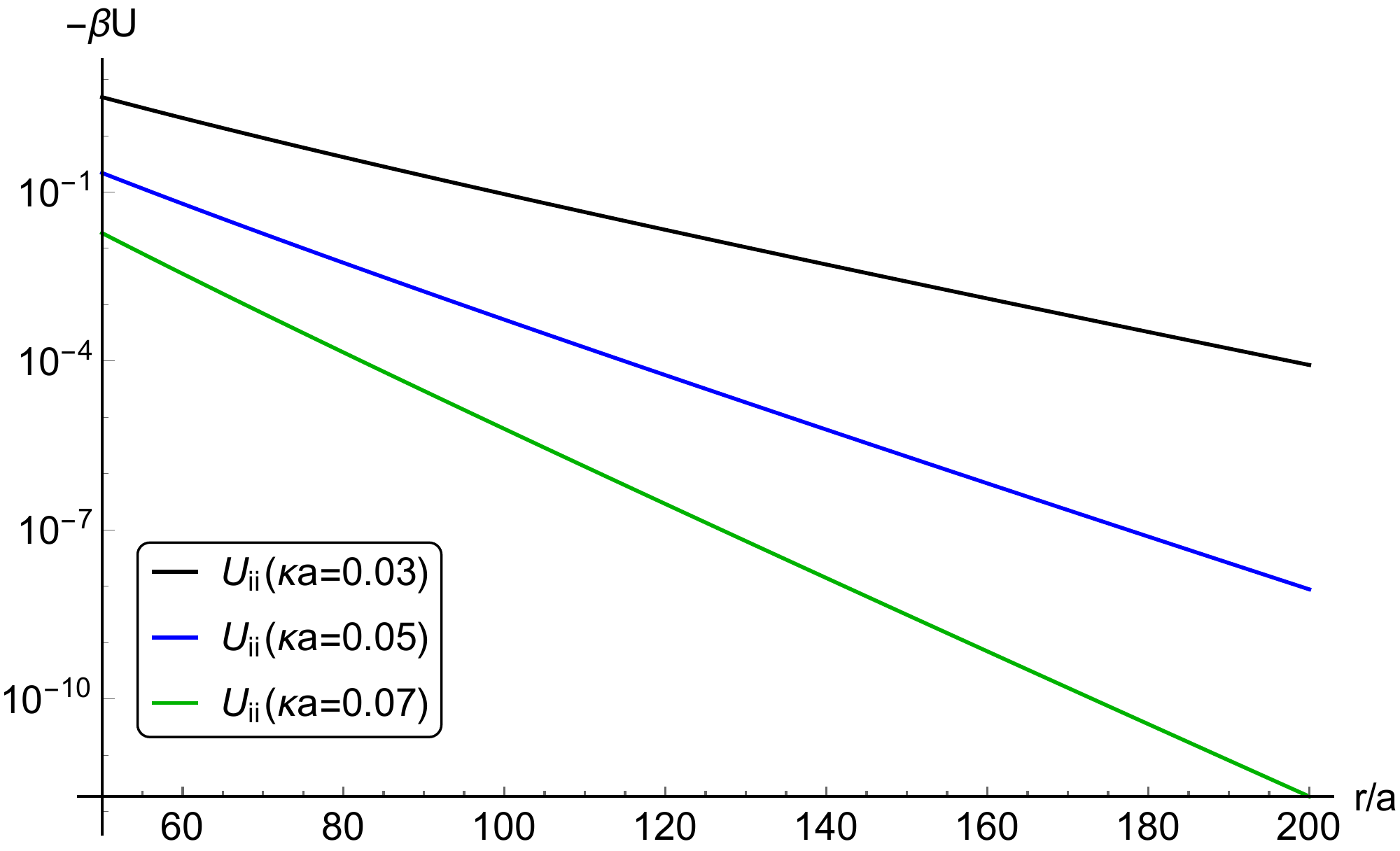}}
\caption{The strength of the ion-image interaction energy, Equation (\ref{ion-image}), is controlled by the (a) metallicity and (b) salt concentration. All variables except the independent variable is set to the typical values estimated in the main text. The ion-image interaction energy is significantly stronger in magnitude compared to the van der Waals energy for distances $r/a \lesssim 240$.  Lower salt concentrations (corresponding to lower $\kappa a$) lessen the electrostatic screening factor and thus are preferred for stronger ion-image interactions. }
\label{vdw_vs_image}
\end{figure}

Figure \ref{vdw_vs_image}a-b shows that the strength of the ion-image interaction energy, Equation (\ref{ion-image}), is controlled by the salt concentration and metallicity. The ion-image interaction is stronger for metallic CNTs with larger metallicity thus higher concentration of free electrons on the CNT surface to respond to the external charge. Increasing the ion concentration decreases the ion-image interaction as counterions screens the electric field from the polyelectrolyte ions. For comparison we also plot an estimate of the van der Waals interaction between a (6,6) CNT and GC(10) dsDNA in water using the \texttt{Gecko Hamaker} software tool \cite{rajter2007van}, showing that the ion-image interaction is significantly larger in magnitude, even at distances much larger than the nanotube size.  At low salt concentrations (long screening lengths) the ion-image interactions become a dominant source of polyelectrolyte-CNT interactions.% for generic polyelectrolytes, barring specific chemical interactions. 

\begin{figure}
\centering
\includegraphics[height=0.3\textwidth]{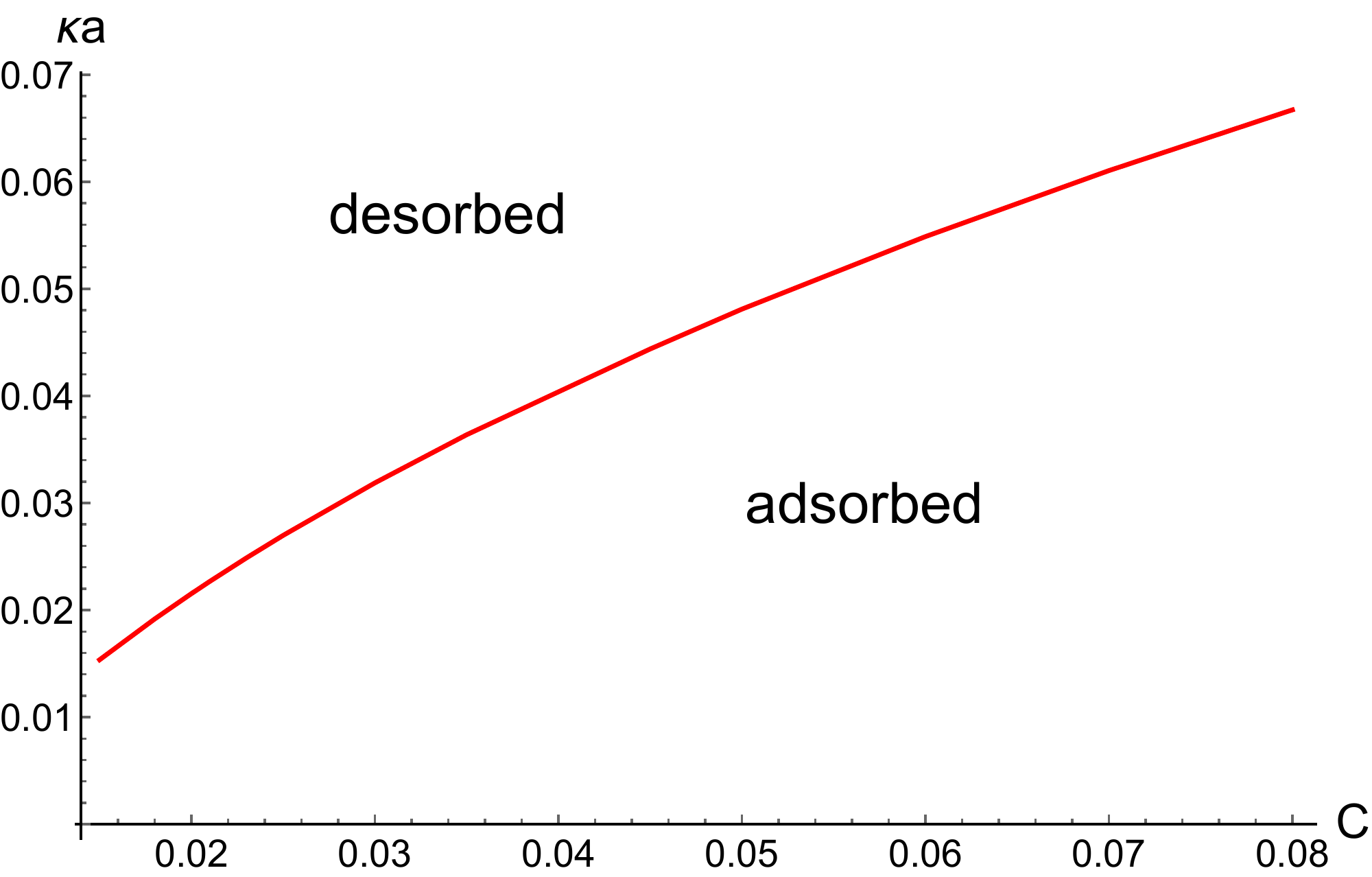}
\caption{Adsorption-desorption phase diagram: Tuning the salt concentration (i.e. $\kappa a$) allows metallicity-selective adsorption of polyelectrolytes onto CNTs. The curve is plotted with the typical parameters estimated in the main text. }
\label{adsorption}
\end{figure}

The dependence of the ion-image interaction on salt concentration and metallicity can be used to control polyelectrolyte adsorption. Figure \ref{adsorption} shows the phase diagram for the adsorption-desorption transition as a function of the metallicity and the inverse Debye length for parameters corresponding to dsDNA. Importantly, tuning the salt concentration allows a metallicity-selective adsorption of polyelectrolytes onto CNTs. At high salt concentration, the ion-image interaction is screened by counterions, thus polymers can only adsorb on CNTs with a large metallicity as the ion-image interaction energy for CNTs with lower metallicity is insufficient to hold the polyelectrolyte at the nanotube surface. 

\begin{figure}
\centering
\includegraphics[height=0.28\textwidth]{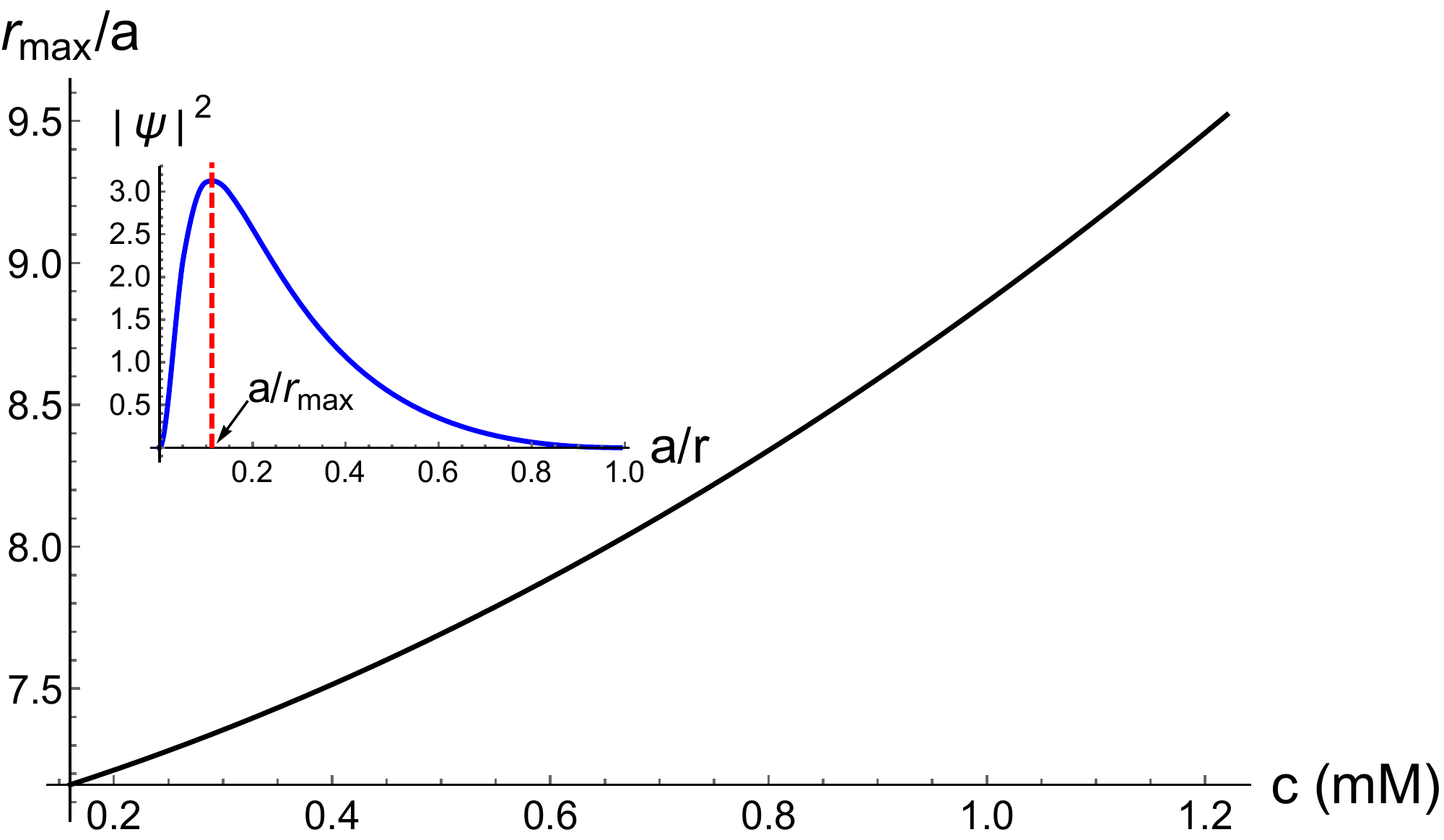}
\caption{The thickness of the adsorbed layer, denoted by $r_{\text{max}}$, increases with salt concentration. Plotted above for a (6,6) CNT with radius $a=0.4068\; \mathrm{nm}$ and $C=C_q/2\pi\varepsilon = 0.05$. Inset: The $c=1\; \mathrm{mM}$ probability distribution of monomers near the CNT surface, which has a maximum at $r_{\text{max}}$.} 
\label{thickness}
\end{figure}

The interplay between Debye screening and ion-image interaction can also be exploited to control the thickness of the adsorbed polyelectrolyte layer. Figure \ref{thickness} shows the thickness of the adsorbed polyelectrolyte layer increases for increasing salt concentration. %, and diverges beyond a critical salt concentration where the polymer layer is desorbed (as shown in the phase diagram Figure (\ref{adsorption})). 
We take the thickness of the adsorbed layer to be the location of the peak in the probability distribution of monomers near the CNT surface (c.f. the inset of Figure (\ref{thickness})). 

In deriving our model, Equations (\ref{eval_problem}), (\ref{ion-image}) and (\ref{OSF}), we have made several assumptions which we will now comment on. First, we considered the weak adsorption regime, where the entropic free energy of polyelectrolyte is comparable to the ion-image interaction with the CNT and thus interactions with the CNT only weakly perturbs the conformation of the chain \cite{cherstvy2006strong}. This is expected to hold near the vicinity of the adsorption-desorption transition, which is the subject of this paper. Second, electrostatic interactions is treated using Debye-H\"{u}ckel theory, which is a mean-field approximation only valid in the asymptotic limit of dilute electrolyte, $l_B^3 c \ll 1$. The adsorption-desorption transition occurs at $O(\mathrm{mM})$, thus $l_B^3 c = O(10^{-4})$ and the mean field approximation is reasonable. Third, we assumed that the polyelectrolyte profile is built up near the adsorbing surface without disturbing the electrostatic potential and ionic distribution near the interface. This approximation is made to render calculations analytically tractable. For constant charge surfaces, a more general self-consistent field theory has been presented in refs. \cite{podgornik1992self,borukhov1999effect}, although a self-consistent computation of just ion-ion interactions near a metallic surface is considerably more complex \cite{hatlo2008role,hatlo2008electrostatic,wang2013effects,wang2016inhomogeneous}, let alone for a polyelectrolyte system. Fourth, we modelled the CNT as an infinite long cylinder. This assumption is justified when the persistence length of the polyelectrolyte is much greater than the persistence length of the CNT (typically $O(100 \; \mathrm{\mu m})$ \cite{fakhri2009diameter}), which holds for the example of double stranded DNA considered here. We have also neglected the rod-like nature of the polyelectrolyte segments in computing the ion-image interaction energy. Finally, the electronic structure of the CNT is modelled using a linearised metallicity (quantum capacitance). For sufficiently large surface potential, the metallicity becomes potential-dependent. In the non-linear regime, the superposition property of linear partial differential equations which we relied on to relate the ion-image interaction energy of a point charge to the ion-image interaction energy of a polyelectrolyte is invalid. 

\section{Discussion and Conclusions}

In summary, we showed how ion-image interaction could drive polyelectrolyte adsorption onto carbon nanotube independent of any chemical affinities between the monomers and the nanotube. The ion-image interaction is tunable by varying the salt concentration and is strongly dependent on the electronic structure of the nanotube. As such, varying the salt concentration is a simple strategy to selectively and reversibly functionalise of carbon nanotubes based on their electronic structure, and the phase diagram mapped by our theory suggests relevant regions of the parameter space.  Experiments and simulations which systematically the effect of metallicity on polyelectrolyte adsorption is scarce; we hope that our theory provides a framework to design experiments and motivate further investigation. In particular, ref \cite{comtet2017nanoscale} reported a set of experimentally viable surfaces with varying metallicity, and extending this pioneering work to investigate polyelectrolyte adsorption would be an exciting direction.

\begin{acknowledgements}
AAL was supported by the George F. Carrier Fellowship at Harvard University and acknowledges the Winton Programme for the Physics of Sustainability at the University of Cambridge for funding. SVK was supported by the Department of Defense (DoD) through the National Defense Science \& Engineering Graduate Fellowship (NDSEG) Program, and by National Science Foundation grant DMS-1411694. MPB is an Investigator of the Simons Foundation.
\end{acknowledgements}

\bibliography{ref}

\end{document}